\documentclass[fleqn,usenatbib]{mnras}

\usepackage[T1]{fontenc}
\usepackage{ae,aecompl}

\usepackage{graphicx}	
\usepackage{amsmath}	
\usepackage{amssymb}	

\usepackage{xcolor}

\hypersetup{draft}

\title{The Random Transiter -- EPIC 249706694/HD 139139}

\author[Rappaport et al.]{
S.~Rappaport$^1$,   
A.~Vanderburg$^{2,3}$, 
M.H.~Kristiansen$^{4,5}$,
M.R.~Omohundro$^{6}$,
\newauthor
H.M. Schwengeler$^{6}$,
I.A.~Terentev$^{6}$,
F.~Dai$^{1,7}$,
K.~Masuda$^{3,7}$,
T.L.~Jacobs$^{8}$,
\newauthor
D.~LaCourse$^{9}$,
D.W.~Latham$^{10}$,
A.~Bieryla$^{10}$,
C.L.~Hedges$^{11}$,
J.~Dittmann$^{12}$,
\newauthor
G.~Barentsen$^{11}$,
W.~Cochran$^{2}$,
M.~Endl$^{2}$,
J.M.~Jenkins$^{13}$.
and A.~Mann$^{14}$,
\\ 
$^{1}$ Department of Physics, and Kavli Institute for Astrophysics and Space Research, M.I.T., Cambridge, MA 02139, USA; sar@mit.edu \\
$^{2}$ Department of Astronomy, The University of Texas at Austin, 2515 Speedway, Stop C1400, Austin, TX 78712 \\
$^{3}$ NASA Sagan Fellow \\
$^{4}$ DTU Space, National Space Institute, Technical University of Denmark, Elektrovej 327, DK-2800 Lyngby, Denmark \\
$^{5}$ Brorfelde Observatory, Observator Gyldenkernes Vej 7, DK-4340 T\o ll\o se, Denmark \\
$^{6}$ Citizen Scientist, c/o Zooniverse, Dept.~of Physics, University of Oxford, Denys Wilkinson Building, Keble Road, Oxford, OX1 3RH, UK \\
$^{7}$ Department of Astrophysical Sciences, Princeton University, Princeton, NJ 08544, USA \\
$^{8}$ Amateur Astronomer, 12812 SE 69th Place Bellevue, WA 98006 \\
$^{9}$ Amateur Astronomer, 7507 52nd Place NE Marysville, WA 98270 \\
$^{10}$ Center for Astrophysics | Harvard \& Smithsonian 60 Garden Street, Cambridge, MA 02138 USA \\ 
$^{11}$ {\em Kepler/K2} Guest Observer Office, NASA Ames Research Center \\
$^{12}$ Department of Earth and Planetary Sciences and Kavli Institute for Astrophysics and Space Research, M.I.T., Cambridge, MA 02139, USA \\
$^{13}$ NASA Ames Research Center, Moffett Field, CA 94035, USA \\
$^{14}$ Department of Physics and Astronomy, University of North Carolina at Chapel Hill, Chapel Hill, NC 27599-3255, USA 
}

\date{}

\pubyear{}

\begin{document}
\label{firstpage}
\pagerange{\pageref{firstpage}--\pageref{lastpage}}
\maketitle

\begin{abstract}
We have identified a star, EPIC 249706694 (HD 139139), that was observed during {\em K2} Campaign 15 with the {\em Kepler} extended mission that appears to exhibit 28 transit-like events over the course of the 87-day observation.  The unusual aspect of these dips, all but two of which have depths of $200 \pm 80$ ppm, is that they exhibit no periodicity, and their arrival times could just as well have been produced by a random number generator.  We show that no more than four of the events can be part of a periodic sequence.  We have done a number of data quality tests to ascertain that these dips are of astrophysical origin, and while we cannot be absolutely certain that this is so, they have all the hallmarks of astrophysical variability on one of two possible host stars (a likely bound pair) in the photometric aperture.  We explore a number of ideas for the origin of these dips, including actual planet transits due to multiple or dust emitting planets, anomalously large TTVs, S- and P-type transits in binary systems, a collection of dust-emitting asteroids, `dipper-star' activity, and short-lived starspots.  All transit scenarios that we have been able to conjure up appear to fail, while the intrinsic stellar variability hypothesis would be novel and untested.

\end{abstract}




\section{Introduction}
\label{sec:intro}

\noindent
The {\em K2} mission \citep{howell14}, spanning four years, has yielded a large number of impressive results regarding what are now considered conventional areas of planetary and stellar astrophysics.  Nearly a thousand new candidate and confirmed exoplanets have been reported (see, e.g., \citealt{mayo18}; \citealt{dattilo19}).  In addition, the {\em K2} mission has continued the {\em Kepler} \citep{koch10} discovery of dynamically interesting triple and quadruple star systems (e.g., \citealt{rappaport17}; \citealt{borkovits19}).  Numerous asteroseismological studies have also been carried out with {\em K2}, including, e.g., \citet{chaplin15}, \citet{lund16a}, \citet{lund16b}, and \citet{stello17}. 

In addition to these areas of investigation, {\em K2} has taken some departures from conventional stellar and planetary astrophysics, and uncovered a number of unexpected treasures.  These include periodic transits by dusty asteroids orbiting the white dwarf WD 1145+017 \citep{vanderburg15}; a disintegrating exoplanet in a 9-hour orbit \citep{sanchisojeda15}; `dipper stars' in the Upper Scorpius Association \citep{ansdell16};  a single day-long 80\% drop in flux in an M star \citep{rappaport19}; numerous long observations of CVs \citep{tovmassian18}; discovery of a hidden population of Blue Stragglers in M67 \citep{leiner19}; a study of asteroids \citep{molnar18}; microlensing events (e.g., \citealt{zhu17}); and continuous before-and-after looks at supernovae and `fast evolving luminous transients' \citep{rest18,dimitriadis19,garnavich19}.

There have also been a number of objects observed during {\em K2} that remain sufficiently enigmatic that they have not yet been published.  In this work, we report on one of these mysterious objects: EPIC 249706694 (HD 139139) that exhibits 28 apparently non-periodic transit-like events during its 87-day observation during {\em K2} Campaign 15.   In Sect.~\ref{sec:search} we report on our continued search for unusual objects in the {\em K2} observations, this time in C15, and the discovery of transit-like dips in EPIC 249706694.  We analyze these dips for periodicities in Sect.~\ref{sec:analyses}, and find none.  In Sect.~\ref{sec:tests} we discuss several tests that we have performed to validate the data set, and, in particular, the observed dips in flux.  In Sect.~\ref{sec:spectra} we discuss the few high-resolution spectra we have acquired of the target star, as well as that of its faint visual companion star.   Various scenarios for producing 28 seemingly randomly occurring transit-like events are explored in Sect.~\ref{sec:scenarios}. We summarize our findings in Sect.~\ref{sec:summary}.

\vspace{0.6cm}

\begin{figure}
\begin{center}
\includegraphics[width=0.995\columnwidth]{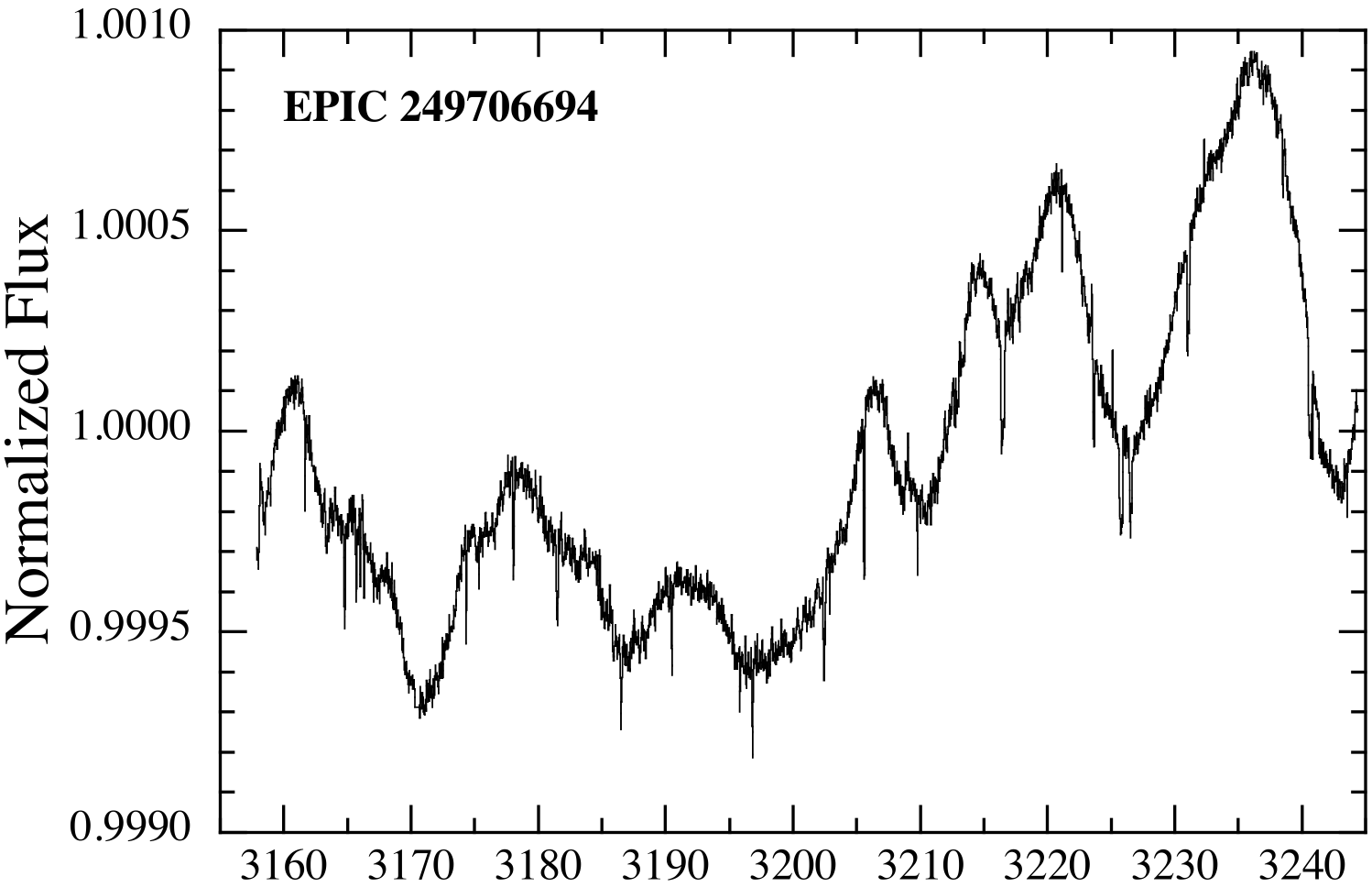} \hglue0.00cm 
\includegraphics[width=1.00\columnwidth]{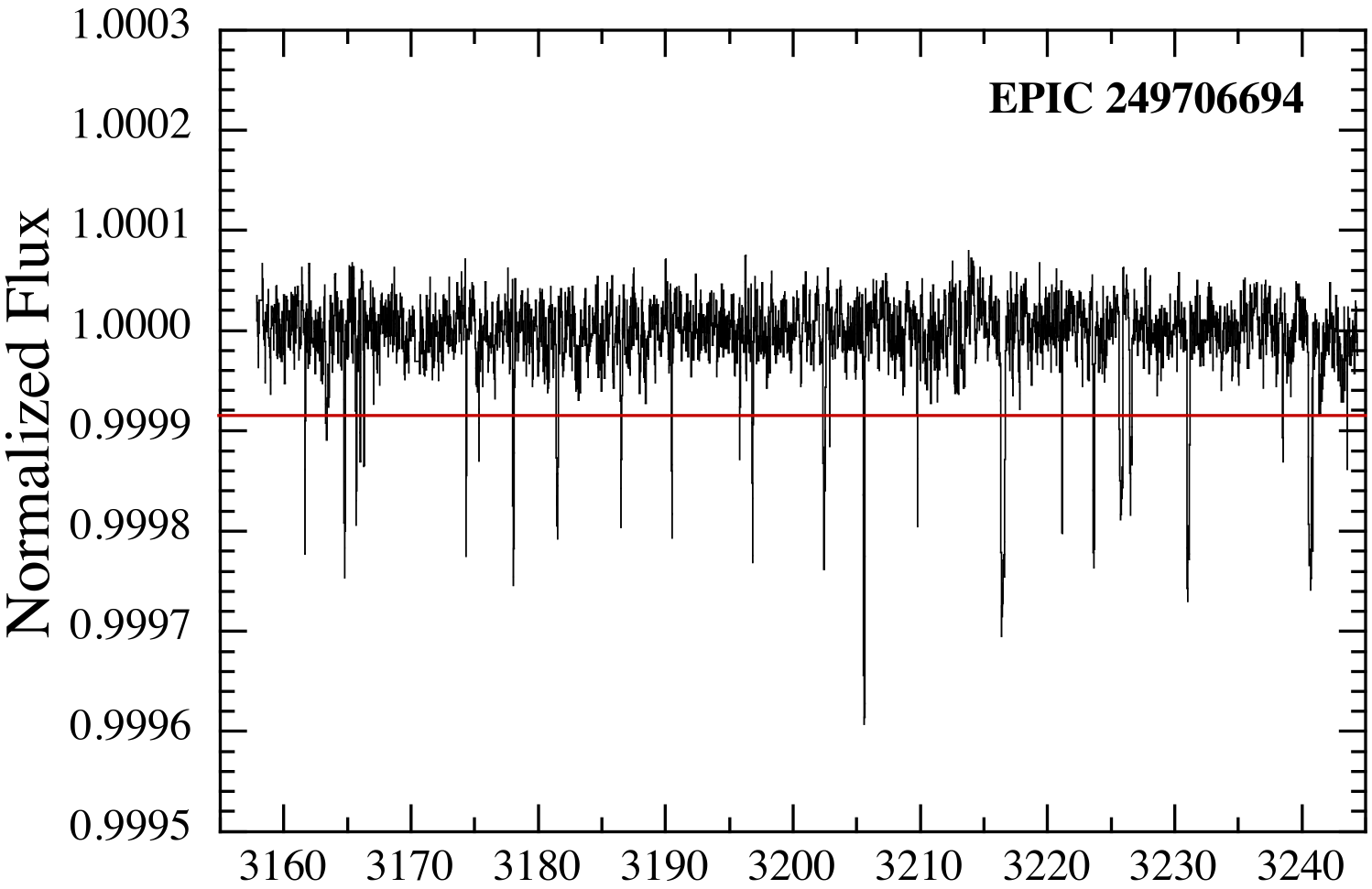} \hglue-0.01cm 
\includegraphics[width=1.005\columnwidth]{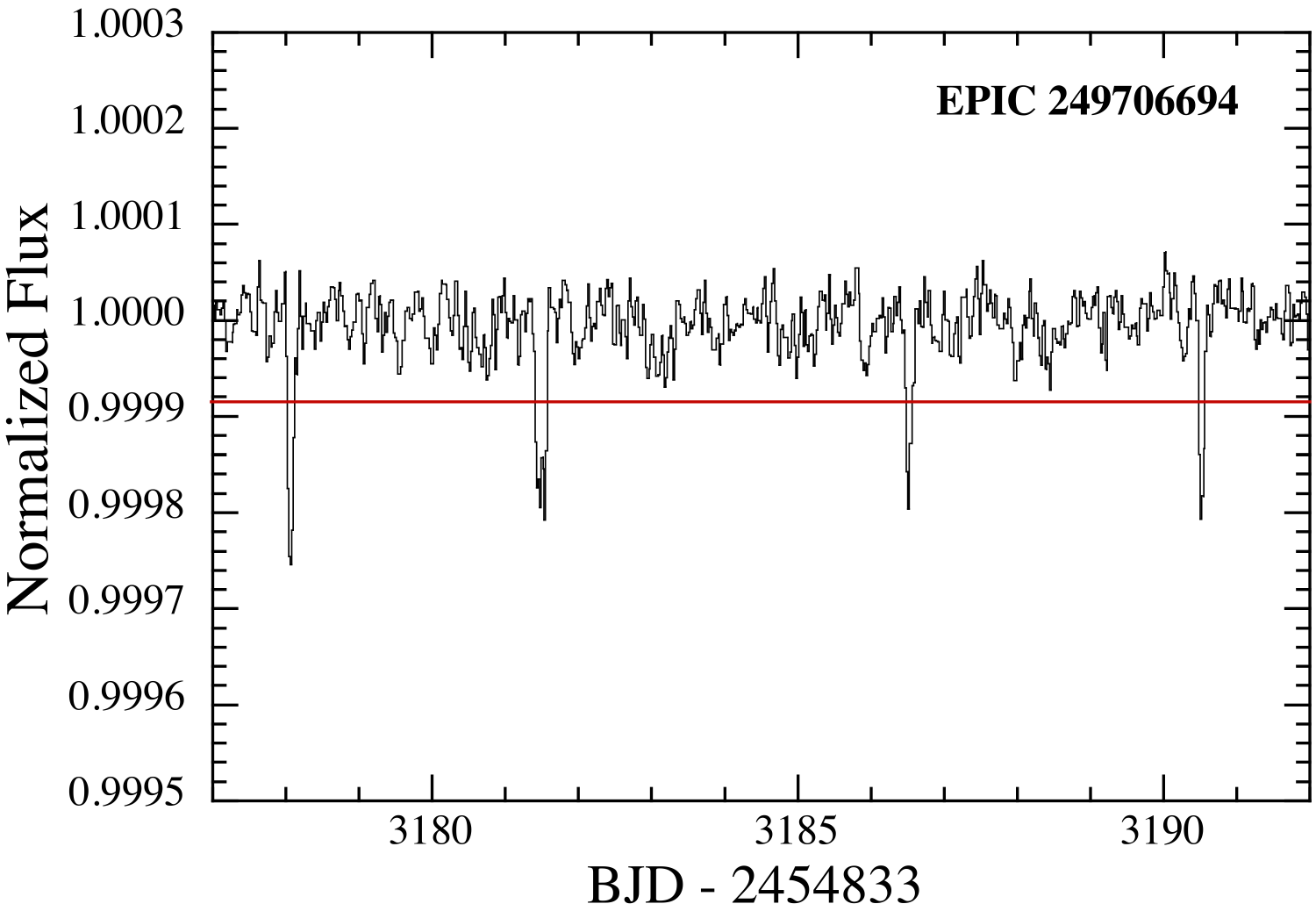}
\caption{{\em K2} lightcurve for EPIC 249706694 (HD 139139) during Campaign 15.  {\em Top panel}: the raw 87-day lightcurve. {\em Middle panel}: lightcurve after filtering out the slow modulations due to star spots and trends that result from data processing. There are 28 transit-like events whose flux dips below the level of the red horizontal line, which we take to be significant (at 82 ppm below unity).  {\em Bottom panel}: a shorter 15-day segment of the lightcurve containing four of the transit-like events.}
\label{fig:rawLC}
\end{center}
\end{figure}  

\section{{\em K2} Data and Results}
\label{sec:K2}

\subsection{{\em K2} Observations and Data}

The {\em K2} Mission (Howell et al. 2014) carried out Campaign 15 (C15) in the constellation Scorpius between 23 August - 20 November 2017, including 23,279 long cadence targets.  Some of the targets overlapped those observed during Campaign 2. At the end of November, the {\em K2} Team made the C15 raw cadence pixel files (`RCPF') publicly available on the Barbara A.~Mikulski Archive for Space Telescopes (MAST)\footnote{\url{http://archive.stsci.edu/k2/data_search/search.php}}.   We first utilized the RCPF in conjunction with the {\tt Kadenza} software package\footnote{\url{https://github.com/KeplerGO/kadenza}}  \citep{barensten18}, combined with custom software, in order to generate minimally corrected lightcurves.  Once the calibrated lightcurves from the {\em Kepler}-pipeline \citep{jenkins10} were released, they were also downloaded from the MAST and surveyed.  Finally, if any interesting objects were found, we utilized the pipelined data set of \citet{vanderburg14} to construct improved lightcurves. 

\subsection{Search for Unusual Objects}
\label{sec:search}

In addition to using Box Least Squares (`BLS'; \citealt{kovacs02}) searches for periodic signals, a number of us (MHK, MO, IT, HMS, TJ, DL) continued our visual survey of the {\em K2} lightcurves searching for unusual objects.  This has led to a number of discoveries not initially caught by BLS searches, including multiply eclipsing hierarchical stellar systems (e.g., EPIC 220204960, \citealt{rappaport17}; EPIC 219217635, \citealt{borkovits18}; EPIC 249432662, \citealt{borkovits19}); multi-planet systems (e.g., WASP 47, \citealt{becker15}; HIP 41378, \citealt{vanderburg16}; EPIC 248435473, \citealt{rodriquez18}); stellar eclipses by accretion disks (EPIC 225300403, \citealt{zhou18}); and a single very deep stellar occultation (80\% deep EPIC 204376071, \citealt{rappaport19}).

Here we present the discovery, via a visual survey, of more than two dozen shallow, transit-like events with durations of 0.75 - 7.4 hours identified in the C15 lightcurve of EPIC 249706694 (aka HD 139139; GO-program PIs: Charbonneau, GO15009; Howard, GO15021; Buzasi, GO15034)\footnote{\url{https://keplerscience.arc.nasa.gov/k2-approved-programs.html\#campaign-15}}. The strange and intriguing thing about these transit-like events is that they exhibit {\em no obvious periodicity}.  In fact, EPIC 249706694 was not flagged by any BLS searches of which we are aware, and only visually spotted while inspecting the lightcurves using LcTools \citep{kipping15}.
   
Though the transit-like events were first spotted in the {\em Kepler}-pipeline lightcurves \citep{jenkins10}, we utilized the data set of \citet{vanderburg14} to construct the lightcurve that is then used throughout the remainder of this study.  The  \citet{vanderburg14} pipeline generates lightcurves starting with pixel-level data.  We subjected this lightcurve to a high-pass filter to remove the starspot activity, and the results are presented in Fig.~\ref{fig:rawLC}.  The top panel shows the data for 87 days of the {\em K2} observation which exhibit the 28 transit-like events that we have detected.  The bottom panel zooms in on a 15-day segment of the data set around four of the transit-like events. 

We tabulate all the depths and durations of the 28 dips in Table \ref{tbl:transits}.  

The robustness of the lightcurve we used was checked by varying both the size and shape of the photometric aperture, and the self flat-fielding (`SFF') parameters, such as the spacing of knots in the spline that models low-frequency variations, the size of the windows on which the one-dimensional flat-fielding correction was performed, and the number of free parameters describing the flat field.  The times, depths, and widths of the dips are quite invariant to these changes.  We also downloaded a lightcurve of EPIC 249706694 generated with the {\tt EVEREST} pipeline \citep{luger16}, and find excellent agreement with our data set.

\begin{figure}
\begin{center}
\includegraphics[width=1.005 \columnwidth]{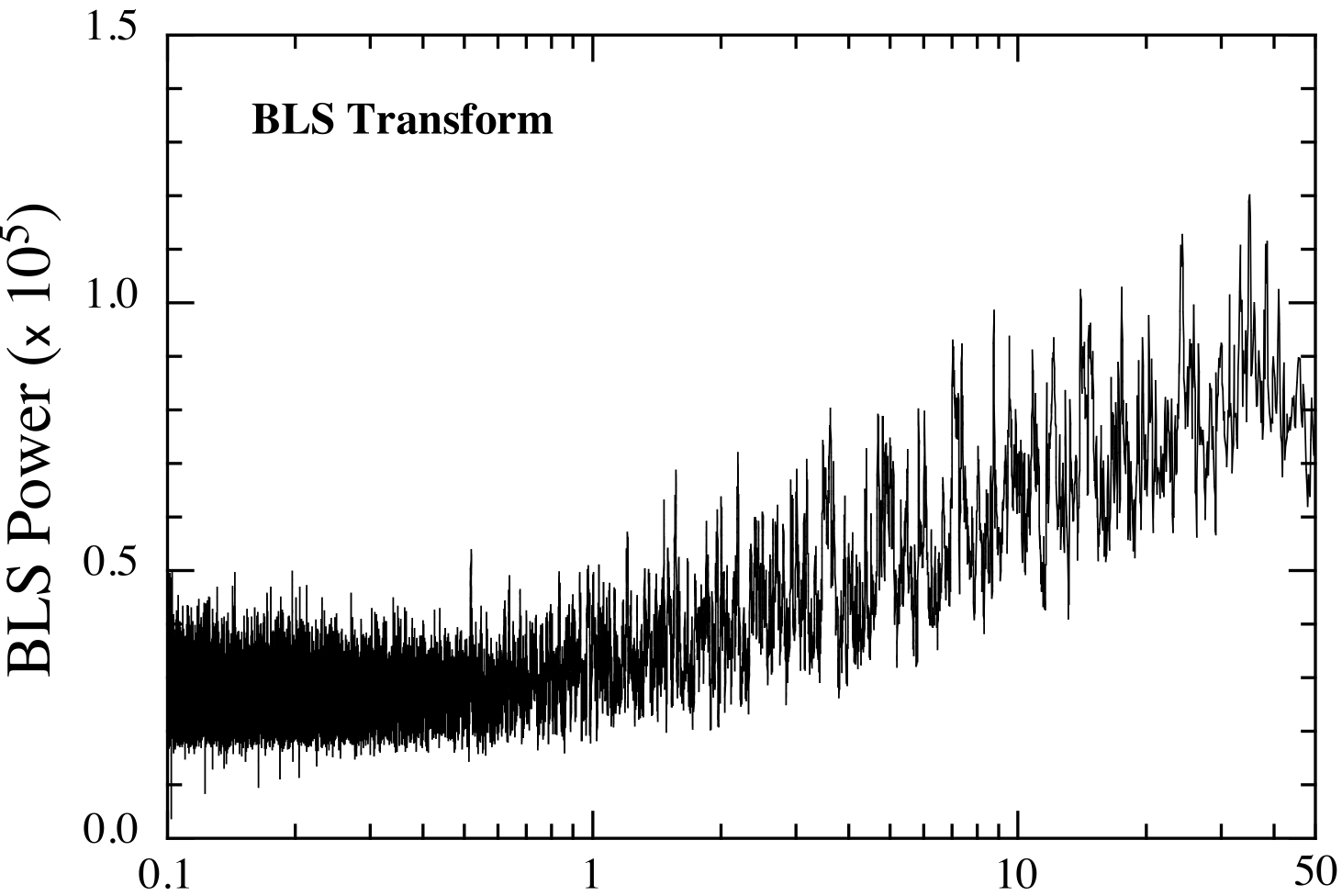} \hglue0.04cm
\includegraphics[width=1.00 \columnwidth]{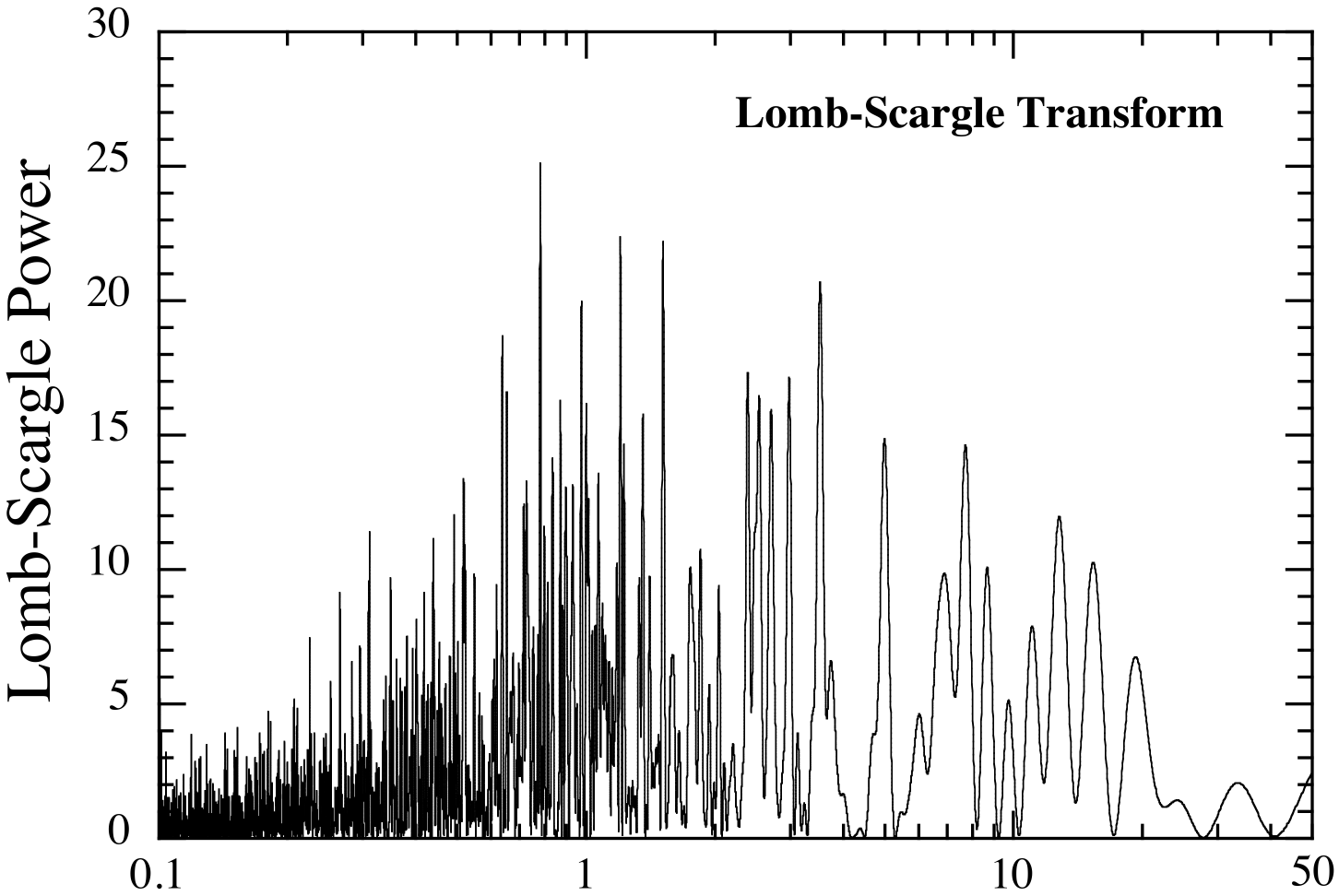} \hglue0.04cm
\includegraphics[width=1.00 \columnwidth]{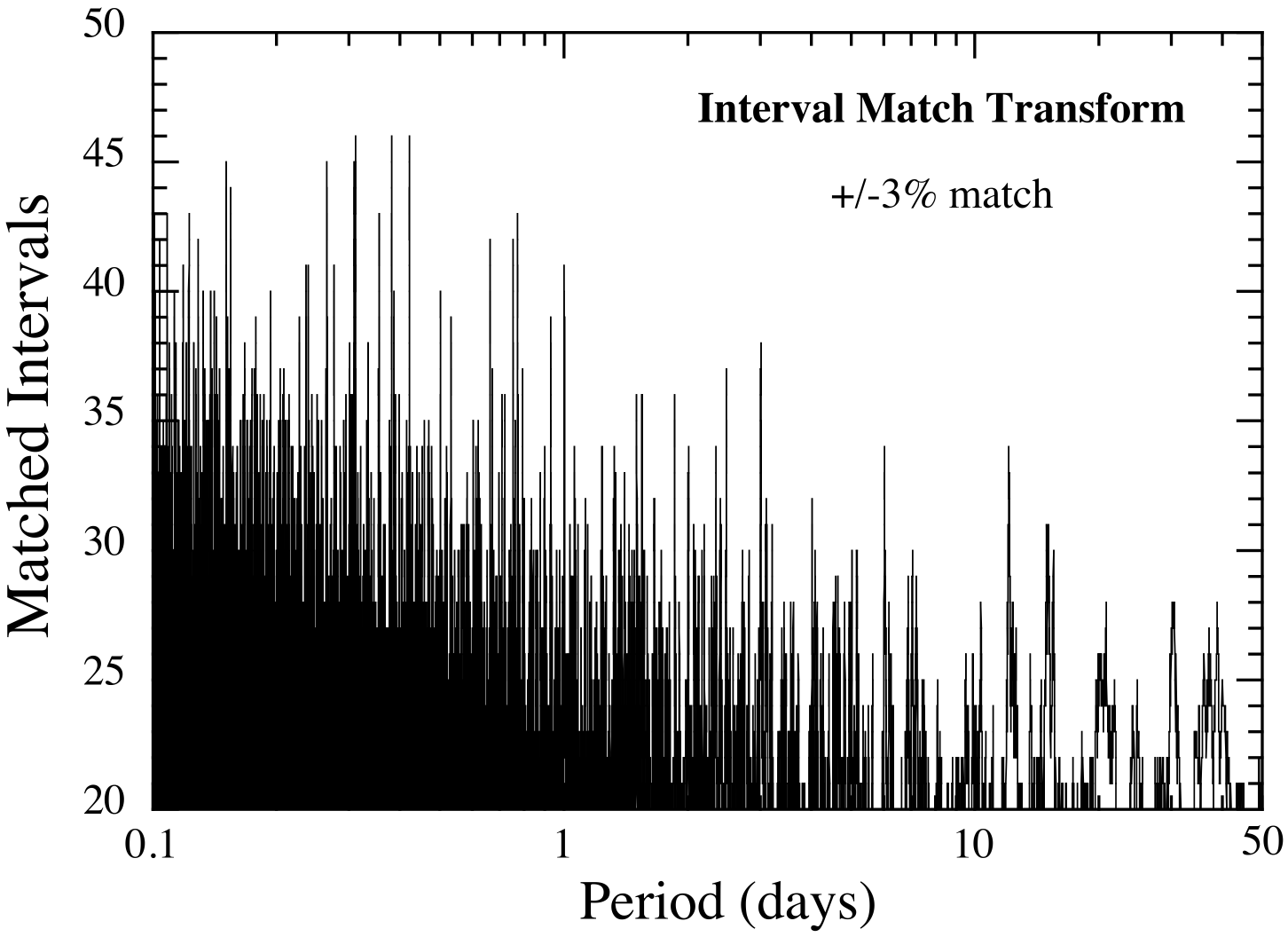}
\caption{Three transforms used to search for periodicities in the 28 transit-like events.  Top panel: Box least squares transform reveals no significant periods in the range of 0.1-50 days.  Middle panel: Lomb-Scargle transform also shows no significant periods over the same range as the BLS.  Bottom panel: An `Interval Match Transform" (see \citealt{gary17}), which allows for very large transit timing variations (`TTVs') of up to $\pm3\%$ of any given trial period, indicates no significant periodicities.  See Sect.~\ref{sec:analyses} for details.}  
\label{fig:transforms}
\end{center}
\end{figure}  

\subsection{Archival Data on EPIC 249706694}
\label{sec:archive}

\begin{figure}
\begin{center}
\includegraphics[width=0.99 \columnwidth]{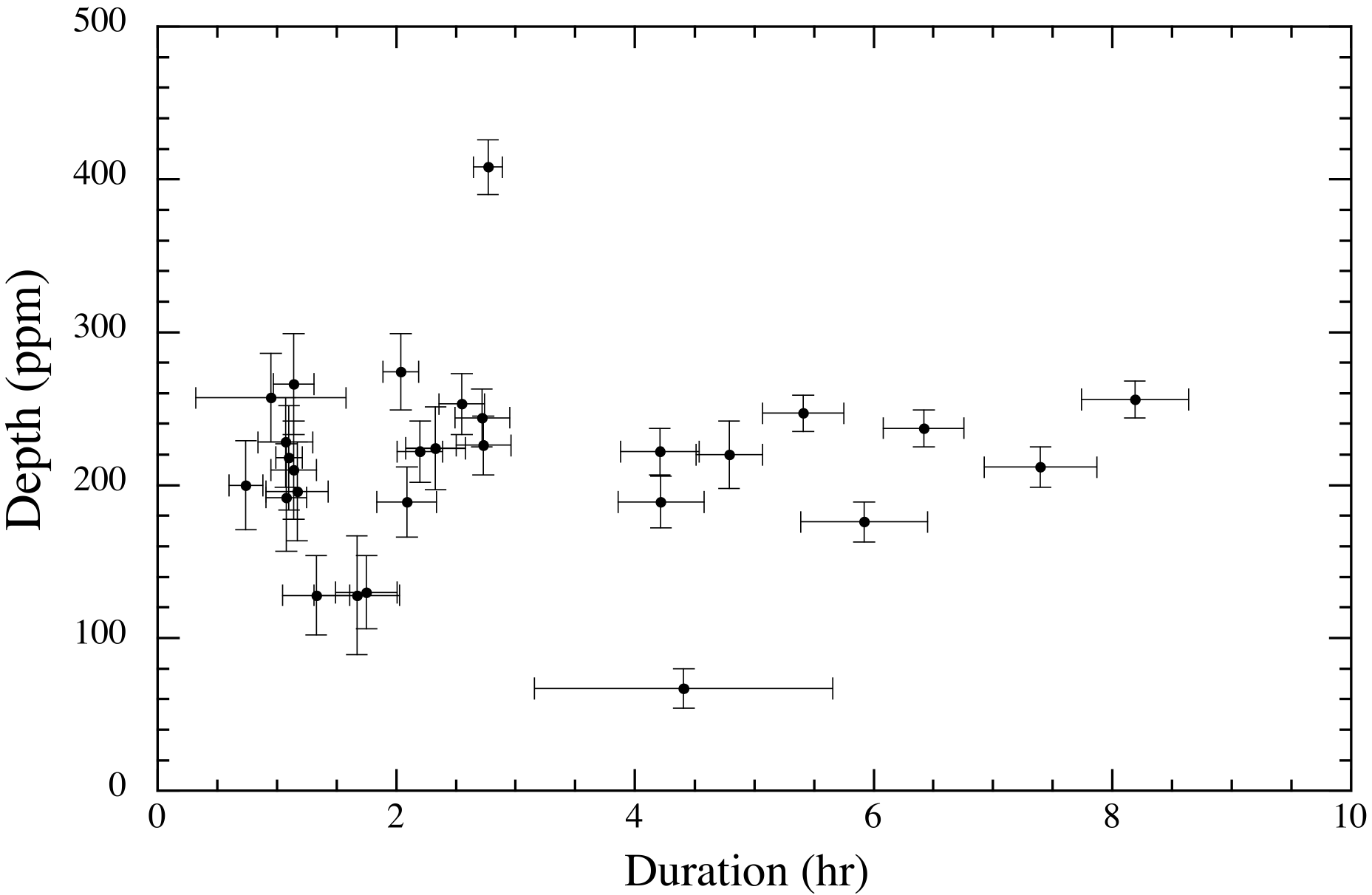}
\caption{The depth (in parts per million) in the flux of the 28 transit-like dips plotted against their duration in hours. There is no apparent correlation between these two parameters which characterize the dips.}
\label{fig:DD_corr}
\end{center}
\end{figure}  

\begin{table}
\centering
\caption{Transit-like Events of the Random Transiter}
\begin{tabular}{lccc}
\hline
\hline
Number & Time & Duration$^a$ & Depth$^a$ \\
  & (BKJD$^b$) & (hours) & (ppm) \\
\hline
1   &    3161.6826(25)   &  1.14(17)   &    266(33) \\ 
2$^c$   &   3163.3811(184)  &   4.41(1.25)  &  67(13) \\ 
3    &   3164.7902(34)   &   2.72(23)  &  244(19) \\ 
4   &   3165.6969(35)    &   2.33(25)   &   224(27) \\ 
5   &    3166.0106(30)   &    1.08(17)   &   192(35) \\ 
6   &    3166.3137(22)   &    1.10(11)  &   218(34) \\ 
7   &    3174.3337(29)   &    2.04(15)  &   274(25) \\ 
8   &    3175.3532(47)   &    1.75(26) &   130(24) \\ 
9   &    3178.0673(29)    &    2.55(19)  &   253(20) \\ 
10   &    3181.4968(53)   &    4.22(36)  &  189(17) \\ 
11    &   3186.5145(43)    &   2.09(25)  &  189(23) \\ 
12   &    3190.5272(30)   &    2.20(19)  &   222(20) \\ 
13   &    3195.8316(48)   &    1.33(28)  &  128(26) \\ 
14   &    3196.8193(34)  &     2.73(23)  &    226(19) \\ 
15   &   3202.4718(42)   &    4.79(28)  &     220(22) \\ 
16   &   3202.9034(49)    &   1.67(36)  &   128(39) \\ 
17   &    3205.6068(18)  &    2.77(12)  &  408(18) \\ 
18   &    3209.7655(30)    &   1.14(19)     &  210(32)  \\ 
19   &   3216.4965(66)   &  8.19(45)    &  256(12)  \\ 
20   &    3221.1631(114)  &   0.95(63)  &   257(29) \\ 
21   &   3223.6633(47)   &    4.21(33)   &   222(15) \\ 
22   &    3225.7971(67)  &    7.40(47)  &   212(13) \\ 
23   &    3226.5548(77)  &    5.92(53)  &  176(13) \\ 
24   &    3231.0569(49)  &    5.41(34)  &    247(12) \\ 
25   &   3238.5381(55)   &    1.17(26)   &   196(32) \\ 
26   &   3240.6550(50)   &    6.42(34)   &   237(12) \\ 
27   &   3240.8371(29)   &    1.07(23)   &  228(29) \\ 
28   &   3243.5688(28)   &    0.74(14)  &   200(29) \\
\hline
\label{tbl:transits}
\end{tabular}  

{\bf Notes.} (a) The depths and durations were determined in two independent ways: (i) fitting to planet transit models and (ii) fitting modified hyperbolic secants (i.e., with a squared argument), and these yielded consistent results. (b) We define ``BKJD'' as BJD-2454833. (c) This dip feature may be suspect because it is the most shallow and a {\em K2} thruster firing occurs within it.  
\end{table}

In Table \ref{tbl:mags} we collect the various archival magnitudes as well as the Gaia kinematic data on EPIC 249706694 (HD 139139).  The target star is quite bright at $K_p = 9.68$ and of solar effective temperature and radius.  However, there is also a neighbouring star (hereafter `star B') that is $\sim$3 magnitudes fainter in $G$, somewhat cooler (4400 K), and separated from the target star by 3.3$''$.  There is almost no additional Gaia information on this star, so we have no initially clear picture of whether star B is a bound companion to star A or not.  

We have done our own astrometry on the elongated 2MASS image (from 1998.4, \citealt{skrutskie06}), the PanSTARRS image (from 2012.1 and saturated in places; \citealt{flewelling16}). and a current-epoch (2019.3) image we acquired at the Fred Lawrence Whipple Observatory (Mt.~Hopkins) using the KeplerCam camera on the 48-inch telescope. Using these data we can typically find the difference in RA and Dec of stars A and B to accuracies of $\sim 0.4'', ~0.1'', ~{\rm and}~0.2''$, for the three images, respectively.  This is sufficiently good, over the $\simeq 21$ year baseline to conclude that the proper motion of star B in both directions tracks that of star A to within $\sim$20 mas yr$^{-1}$.  Our astrometry is summarized in Table \ref{tbl:astrometry}. Given that the proper motions in RA and Dec of star A are $-68$ and $-92$ mas yr$^{-1}$, respectively, this boosts the likelihood that these two stars are a bound pair.

Additionally, we attempted to estimate a photometric distance to star B. We assume for simplicity that star B (having a low $T_{\rm eff}$) is still on the main sequence, has solar metallicity, and has a radius appropriate to its $T_{\rm eff}$. We matched the photometry to optical templates from \citet{gaidos14}, which implied a spectral type of K5-K7, a temperature of 4100-4300\,K, and an absolute Gaia-$G$ magnitude ($M_G$) of 7.2-7.6 (all ranges are 1$\sigma$). Using the resolved Gaia $G$ photometry of B, this implies a distance of 115$\pm$20\,pc. This is consistent with the Gaia distance to A (106.7-107.8\,pc; \citealt{bailerjones18}) and further enhances the likelihood that stars A and B form a bound pair.  

There are both ASAS-SN (5 years; e.g., \citealt{jayasinghe18}) and DASCH (100 years; \citealt{grindlay09}) lightcurves for the target star.  Neither of these lightcurves indicates any particularly interesting activity, i.e., outbursts, dimmings, or periodicities, and therefore they are not shown here. 

For convenience, we note that EPIC 249706694 was not targeted in C2 nor is there an observation window scheduled for {\em TESS} \citep{ricker14} based on the Web TESS Viewing Tool\footnote{\url{https://heasarc.gsfc.nasa.gov/docs/tess/data-access.html}}.

\begin{table}
\centering
\caption{Properties of the EPIC 249706694 System}
\begin{tabular}{lcc}
\hline
\hline
Parameter & Star A & Star B \\
\hline
RA (J2000) & 234.27559 & 234.27632 \\  
Dec (J2000) &  $-$19.14292 & $-$19.14353 \\  
$K_p$ & 9.677 & ... \\
$G$$^a$ & $9.562 \pm 0.0002$ & $12.684 \pm 0.001$ \\
$G_{\rm BP}$$^a$ & $9.907 \pm 0.001$ & $ 13.320 \pm 0.060 $ \\
$G_{\rm RP}$$^a$ & $9.086 \pm 0.001$ & $11.762 \pm 0.008 $ \\
J$^b$ & $8.435 \pm 0.030$  & ... \\
H$^b$ & $8.101 \pm 0.040$ & ... \\
K$^b$ & $8.050 \pm 0.030$ & ... \\
W1$^c$ & $7.916 \pm 0.024$ & ... \\
W2$^c$ & $7.970 \pm 0.020$ & ... \\
W3$^c$ & $7.984 \pm 0.021$ & ... \\
W4$^c$ & $7.828 \pm 0.219$ & ... \\
$T_{\rm eff}$ (K)$^a$ & $5765 \pm 100$ & $4407 \pm 130$ \\
$R$ ($R_\odot$) & $1.14^{+0.03}_{-0.05} $ & ... \\
$L$ ($L_\odot$) & $1.29 \pm 0.01 $ & ... \\
Rotation Period (d)$^d$ & $14.5 \pm 0.7$ & ... \\
Distance (pc)$^a$ & $ 107.6 \pm 0.5$ & ... \\   
$\mu_\alpha$ (mas ~${\rm yr}^{-1}$)$^a$ & $-67.6 \pm 0.09$  & e \\ 
$\mu_\delta$ (mas ~${\rm yr}^{-1}$)$^a$ &  $-92.5 \pm 0.06$  & e \\ 
\hline
\label{tbl:mags}  
\end{tabular}

{\bf Notes.} Stars A and B are separated by 3.3$''$. (a) Gaia DR2 \citep{lindegren18}.  (b) 2MASS catalog \citep{skrutskie06}.  (c) WISE point source catalog \citep{cutri13}.  (d) Derived from an autocorrelation function made with the raw {\em K2} lightcurve. (e) Based on our own astrometry of a 2MASS, PanSTARRS, and recent FLWO image using KeplerCam, we find that the proper motion is the same as for star A to within $\sim$20 mas yr$^{-1}$ in RA and in Dec (see Sect.~\ref{sec:archive}). 
\end{table}
   
\section{Analysis of the Dips}
\label{sec:analyses}

We have searched extensively for periodicities among the 28 transit-like events.  In the top panel of Fig.~\ref{fig:transforms} we show the box least-squares spectrum. There are no significant peaks.  The middle panel of Fig.~\ref{fig:transforms} is the result of a Lomb-Scargle transform, and, as can be expected if the BLS shows no signal for transit-like events, neither will an LS transform.  

In the bottom panel of Fig.~\ref{fig:transforms} we show the results of an `Interval Match Transform' (`IMT'; see Sect.~7 of \citealt{gary17}) which consists of the brute force testing of $10^5$ trial periods against all unique ($28 \times 27/2 = 378$) dip arrival-time differences.  For each trial period we checked whether the time interval between each distinct pair of dips was equal to an integer number of the trial period.  We allowed for a plus or minus 3\% leeway in terms of matching an integral number of cycles. (This tolerance for matching an integer can be set arbitrarily to allow for small or even large TTVs.)  The IMT in this case should produce a $\sim$6\% accidental rate of matches for periods that are (i) long compared to the shortest inter-dip interval ($\sim$0.2 d) and (ii) short compared to the length of the data train (i.e., 1-10 days), yielding $\sim$$0.06 \times 378$ = 23 accidental matches.  Therefore, if even 1/4 of the 28 events are truly periodic, then there should be an additional 21 (i.e., $7 \times 6/2)$ consequential matches at the true periodicity.  And, we see no such indication.

From these tests we know that not more than about 7 of the 28 dips in flux can be periodic.  But, perhaps there are small-number sets of periodic dips.  In order to test just how sensitive the BLS algorithm is to a limited set of periodic dips in the presence of numerous non-periodic dips, we did the following exercise.  First we injected a set of 10 periodic transits of depth 200 ppm and duration 3 hours into the actual data set containing the 28 apparently random transits.  This choice of depth and duration seems reasonably typical of the other dips.  Those 10 periodic transits were easily detected at high signal-to-noise with the BLS search.  We then repeated the exercise, but with 9, 8, 7, ... etc.~added periodic dips.  They were readily detectable down to the addition of only 4 periodic dips.  Three dips (and obviously any fewer) could not be convincingly detected.  In Table \ref{tbl:BLS} we list the number of added periodic transits to the data set, and the signal-to-noise with which they were detected.  From this we conclude that any periodicities among the 28 dips in flux can occur in sets no larger than four events each. 

In Fig.~\ref{fig:DD_corr} we show a plot of the dip depth versus their duration.  As is fairly obvious, there is no correlation between the two parameters.  All but two depths lie in the range $200 \pm 80$ ppm.  

\begin{figure}
\begin{center}
\includegraphics[width=1.00 \columnwidth]{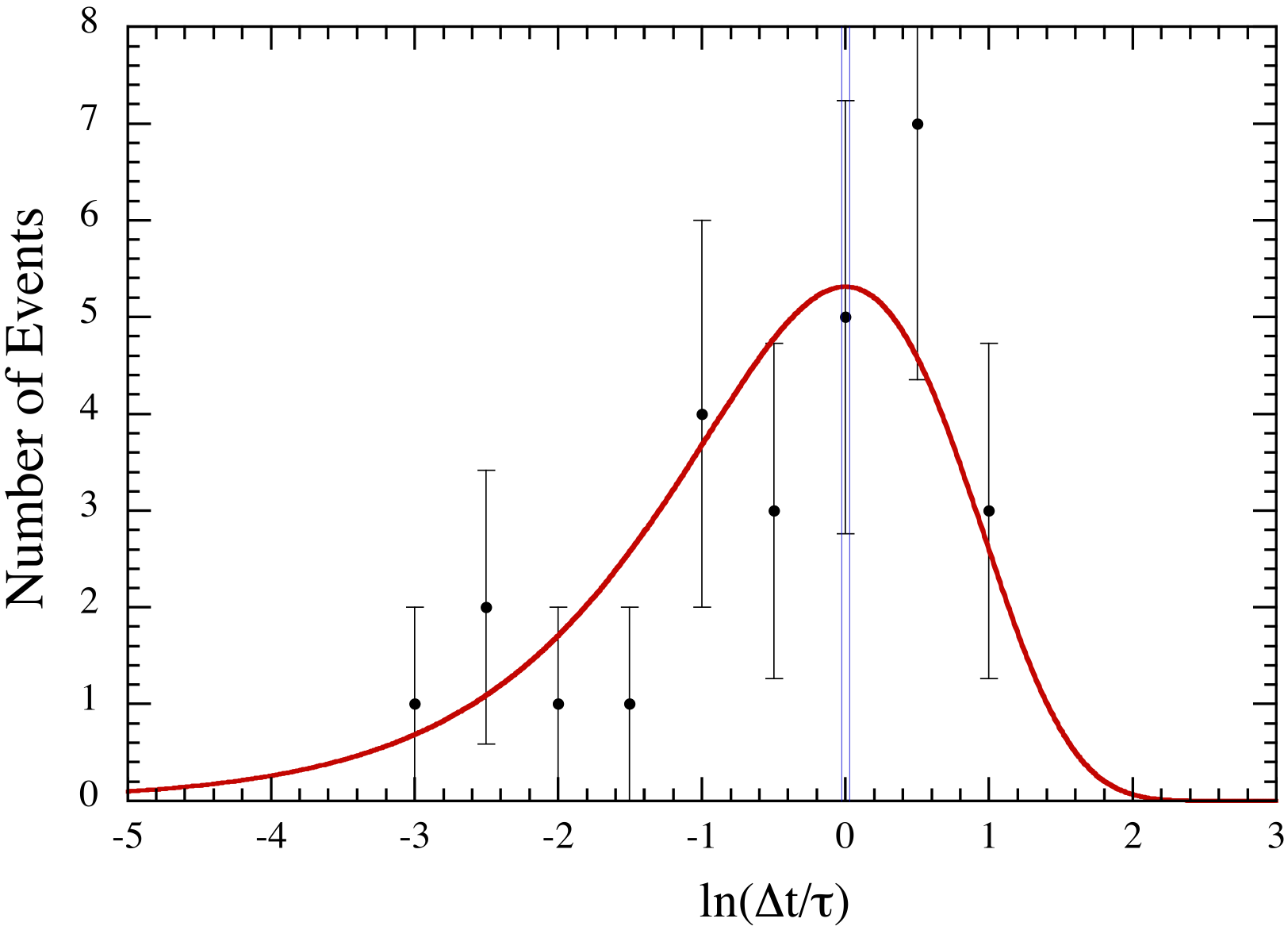}  
\caption{Distribution of the sequential interarrival times, $\Delta t$, of the 28 dips, expressed as the natural log of $\Delta t/\tau$, where $\tau$ is the mean rate of transit-like events (3.09 per day).  The red fitted curve is the `exponential' distribution, expressed in units of per logarithmic interval, that would be expected for random arrival times.  The narrowly spaced vertical blue lines are what would be expected for a periodic function with 3\% TTVs.}
\label{fig:intertime}
\end{center}
\end{figure}

We did one formal test for random arrival times among the 28 observed dips in flux.  For a Poisson distribution of arrival times, the `survival' probability for having no event following a given event for a time $\Delta t$ is proportional to $\exp(-\Delta t/\tau)$, where $\tau$ is the mean time between arrivals (3.1 days in our case).  Since there are relatively few events to work with, and the values of $\Delta t$ cover two orders of magnitude, we decided to use the logarithm of the $\Delta t$'s.  The distribution of arrival time differences is then given by 
\begin{equation}
dN/dz = N \exp[z-\exp(z)]
\label{eqn:poisson}
\end{equation}
where $z \equiv \ln(\Delta t/\tau)$, and $N$ is the number of arrival time differences.  The distribution of arrival times (in logarithmic bins) and a fit to Eqn.~(\ref{eqn:poisson}) is shown in Fig.~\ref{fig:intertime}.  The only free parameter is the normalization, and that was computed using Cash statistics \citep{cash79} for the small numbers in each bin. The fit is good, considering the limited statistics.  In the end, with only 28 events it is difficult to prove formally that a data set actually has random arrival times.  And, it is likely that the arrival times of a several-planet system would pass this same test (i.e., a fit to Eqn.~\ref{eqn:poisson}).

\begin{figure}
\begin{center}
\includegraphics[width=1.00 \columnwidth]{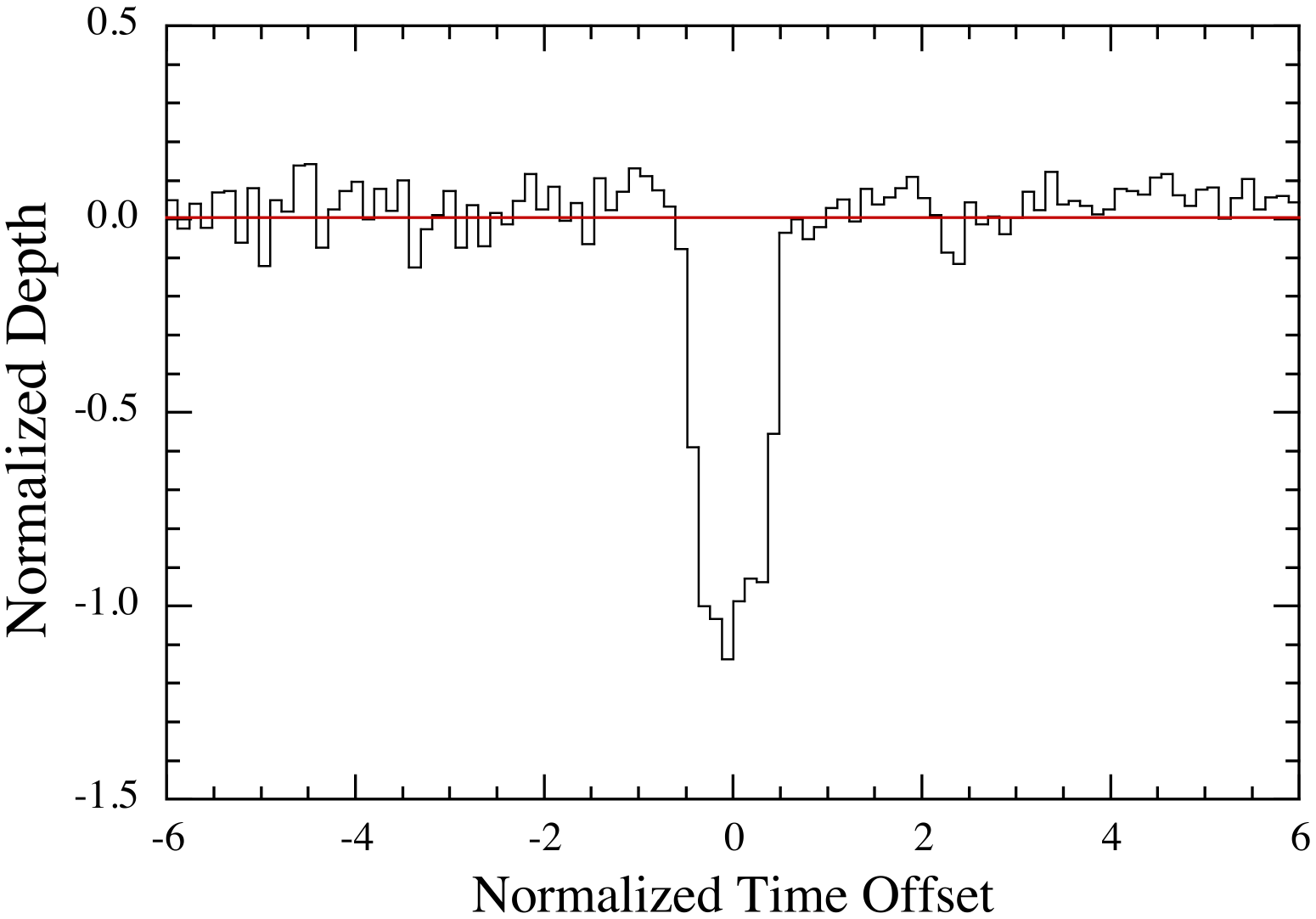}  
\caption{Average profile of the 28 detected dips in EPIC 249706694 (HD 139139). The dips are shifted to a common center, are stretched to a common duration of 2 hours, and expanded vertically to a common depth of 200 ppm. One unit on the X and Y axes corresponds to 2 hours and 200 ppm, respectively.  No significant asymmetries, long egress tail, or pre- or post-transit features are seen.}
\label{fig:stack}
\end{center}
\end{figure}

Finally, we created an average profile for the 28 dips by stacking them (see Fig.~\ref{fig:stack}). Each dip is shifted to a common center, and is then stretched (or shrunk) to a common duration of 2 hours, and expanded (or contracted) vertically to a common depth of 200 ppm. It is not surprising, therefore, that the average profile has a duration close to 2 hours and a depth near 200 ppm.  However, what is more significant is that the average profile is closer to a planet-transit shape than to being triangular, and that no statistically significant asymmetries, long egress tail, or pre- or post-transit features are seen (e.g., as in the cases of the so-called `disintegrating planets', \citealt{vanlieshout18}.)

\begin{table}
\centering
\caption{Relative Astrometry: Star A vs.~Star B}
\begin{tabular}{cccc}
\hline
Image source & Effective Date & $\delta$RA & $\delta$Dec \\
         & JD - 2450000  & arc sec  &  arc sec \\
\hline
2MASS$^a$ & 0975 & $2.36 \pm 0.39$ & $2.51 \pm 0.40$ \\
PanSTARRS$^a$ & 5967 & $2.29 \pm 0.11$ & $2.12 \pm 0.11$ \\  
Gaia & $\sim$7212 & $2.48 \pm 0.00$ & $2.18 \pm 0.00$ \\
KepCam$^a$ & 8602 & $2.47 \pm0.19$ & $2.03 \pm 0.18$ \\
\hline
\label{tbl:astrometry}
\end{tabular}  

{{\bf Notes.} (a) See Sect.~\ref{sec:archive} for details.} 
\end{table}

\section{Tests of the Data Set}
\label{sec:tests}

It is possible for the instrument systematics of the {\em Kepler} detector system to mimic planetary transits. In particular, the `rolling band artefact' can cause small, transit-shaped changes in flux (see the {\em Kepler} Instrument Handbook; \citealt{thompson16}), and channel ``cross-talk'' can produce spurious signals. Additionally, transiting signals originating from background targets could pollute the target and mimic transits in EPIC 249706694. We have vetted EPIC 249706694 against these common types of instrument systematic and performed a light centroid test to check for background contaminants, and we find no evidence that the transiting signals are caused by instrument systematics or background stars.

Based on our analysis, the signal is unlikely to be due to rolling band artefacts for the following reasons.  First, rolling band events are usually infrequent, and low signal-to-noise, unlike the transits in EPIC 249706694. Second, since rolling band is an additive effect, background pixels are proportionally more affected by rolling band, and the background pixels of EPIC 249706694 do not show any transiting signals.  Third, we note that no rolling bands have been detected on the particular channel containing our target star \citep{vancleve16} so that there is little risk that the features we are observing are due to rolling bands. However, out of an abundance of caution, we have also analyzed the lightcurves of the 10 nearest neighbours to EPIC 249706694 and find no evidence of transits in either their lightcurves or background.  These neighbour stars are listed in Table \ref{tbl:neighbours}. 

EPIC 249706694 is a bright target, and in order to create a spurious background signal of 200 ppm that would mimic a transit would require a significant contribution from an additive background. We find that all ten of the nearest neighbours of EPIC 249706694 would have exhibited the equivalent, additive background signal at greater than 8 $\sigma$ confidence. As we do not detect any transits in background apertures or neighbouring stars, we rule out an additive, rolling band type signal as the cause of these dips.

In addition to the {\em Kepler} magnitudes of the neighbouring stars, Table \ref{tbl:neighbours} also gives the limiting depth (2-$\sigma$) that could have been detected for dips in flux lasting for 3 hours, were they to host a multiplicative, contaminating signal.  Seven of the stars are faint, ($K_p$ between 14.9 and 18.6), and we would not have achieved the sensitivity to detect the dips in EPIC 249706694. For three of the neighbours with $K_p$ of 10.7 to 12.9, we would have the sensitivity to detect 200 ppm dips lasting for 3 hours; however we detect no such dips in their lightcurves. This further suggests that the signal is not an instrument systematic.

It is possible to cause a spurious signal from detector ``cross talk'', in which bright targets on one channel cause excess flux to be measured on other channels in the same module, at the same row/column coordinates.  However, cross talk is a small effect, and requires a bright target on a channel in the same module. To create a spurious signal of $\sim$500 counts to cause these transiting signals would require a bright target on adjacent neighbouring channels. In this case, as EPIC 249706694 is on channel 42, to create the transiting signal would require a star of 10th magnitude on channels 41 or 43, or 6th magnitude on channel 44.  Using the {\em K2} Campaign 15 full-frame images, we find that there is no target bright enough to cause cross talk on these channels.  

We have checked the Pixel Response Function (PRF) centroids for EPIC 249706694 during the dips, as compared to out-of-dip regions. PRF centroid shifts during transits would indicate that the transiting signal is coming from a neighbouring or background target. We are able to establish that the PRF centroid during the transit is consistent with the PRF centroid out of transit, suggesting that the transiting signal originates from EPIC 249706694. To strengthen this interpretation, given that this star is saturated and bleeding, we formulated two sets of difference images of the average out-of-transit image versus the average in-transit image, where the out-of-transit data were taken to be one transit-duration on either side of each transit. In one set we fitted out the measured motion of the star from each pixel prior to constructing the difference images to reduce the impact of the pointing deviations. In both cases we found that many of the difference images exhibited features associated with the transits at the ends of the saturated bleed segments. In other cases the difference image showed that the transit-like feature was occurring in the core of the stellar image on either the left or the right side of the saturated, bleeding columns. These difference images indicate that the source of the dips is collocated with, or very close to, the target star.

As reported in Table \ref{tbl:mags}, a neighbour star exists within 3.3$''$ of EPIC 249706694.  Since this star exists in the bleed column of EPIC 249706694, we are not able to discount it as the origin of the transiting signals.

\begin{table}
\centering
\caption{BLS SensitivityTests$^a$}
\begin{tabular}{ccc}
\hline
Injected transits & Period (d) & S/N \\
\hline
10 & 8.6 & 18.0 \\
9 & 9.4 & 17.3 \\
8 & 10.9 & 13.9 \\
7 & 12.4 & 12.6 \\
6 & 14.0 & 10.9 \\
5 & 16.9 & 9.2 \\
4 & 22.0 & 7.2 \\
3 & 30.2 & 4.0 \\
\hline
\label{tbl:BLS}
\end{tabular}  

{{\bf Notes.} (a) Detection sensitivity to small numbers of periodic transits in the presence of 28 non-periodic ones.} 
\end{table}

\begin{table}
\centering
\caption{Nearest Neighbour$^a$ Stars to EPIC 249706694}
\begin{tabular}{lccc}
\hline
Star & $K_p$ & Limiting$^b$ 3-hr Dip & Separation$^c$ \\
(EPIC) &   & (ppm)  & arc min \\
\hline
249704360 & 12.9  & 94  & 1.8 \\
249708809 & 17.9 & 1372 & 2.2 \\
249705070 & 10.7 & 32 & 2.8 \\
249702101 & 17.8 & 1100 & 3.8 \\ 
249704543 & 16.9 & 1586 & 4.4 \\
249710690 & 16.7 & 1401 & 5.3\\  
249700088 & 12.5 & 76 & 5.3 \\  
249713491 & 14.9 & 281 & 6.2 \\   	
249713323 & 18.6 & 2014 & 6.2 \\   	
249713897 & 17.9 & 4179 & 6.3 \\ 
\hline
\label{tbl:neighbours}
\end{tabular}  

{\bf Notes.} (a) All the {\em K2} observed stars within 6$'$ of EPIC 249706694. (b) 2-$\sigma$ sensitivity limit. (c) Angular separation between EPIC 249706694 and the neighbour star.
\end{table}

\begin{figure}
\begin{center}
\includegraphics[width=1.00 \columnwidth]{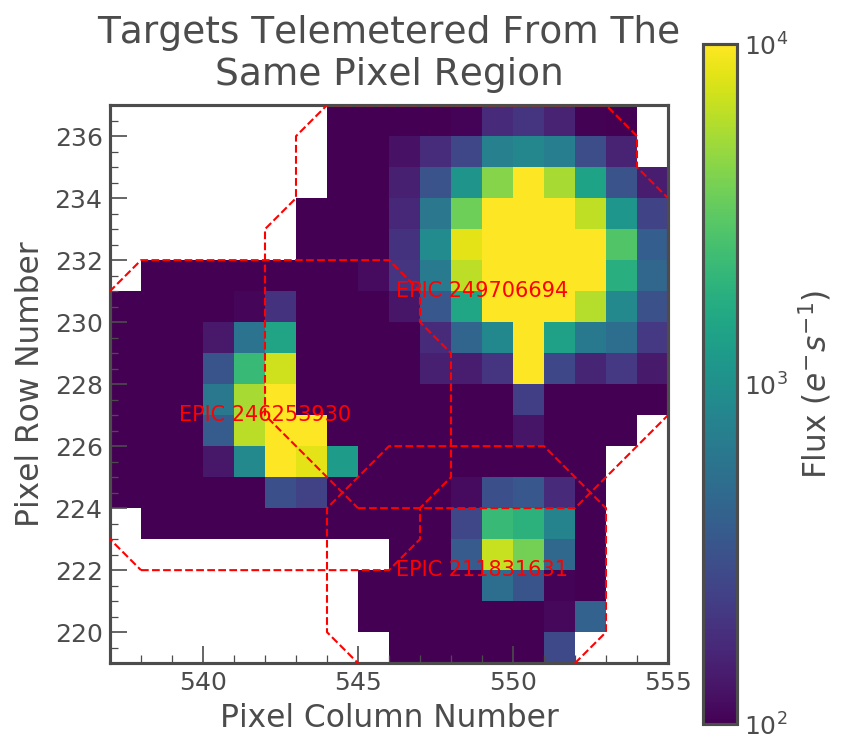}  
\caption{Composite image of EPIC 249706694 showing our target star as well as two other EPIC stars 211831631 and 246253930, each observed during a different campaign.  The red lines denote the postage-stamp boundaries for each star during its respective observation.  We have verified that during observations of EPIC 211831631 and 246253930  the pixels in the overlap region were behaving normally.}
\label{fig:pixels}
\end{center}
\end{figure}  

We have also checked the performance of the CCD pixels which detected the transit-like signals in EPIC 249706694 during observations of other stars during previous {\em K2} campaigns.  Figure \ref{fig:pixels} shows a composite image of EPICs 249706694, 211831631, and 246253930 (the latter two from C5 and C12, respectively). Based on the measurements of these pixels from other campaigns, we have no evidence that there is a detector anomaly causing these transit-like signals.  However, the overlap in pixels is only partial, and so we cannot formally exclude an anomaly in a group of pixels that do not overlap the nearby C5 and C12 targets (Fig.~\ref{fig:pixels}).  Known detector problems include the `Sudden Pixel Sensitivity Drop Out' (`SPSD'; see, e.g., \citealt{christiansen13}) which causes permanent degradation, and would not lead to the transit-like behavior we observe.  Although we cannot completely rule out a detector anomaly as the source of the signal, we have no evidence to suggest that the pixels used to detect EPIC 249706694 are faulty, and no other reported detector anomaly shows a similar morphology to the transits we observe.

Finally, we have also considered the possibility that solar system objects (e.g., asteroids) passing through or near the photometric aperture could somehow cause spurious dips in flux. We used {\tt SkyBot} \citep{berthier6,berthier16} and find that at the arrival times of the dips (Table \ref{tbl:transits}) there are no known catalogued asteroids within 1$'$ of the target star (EPIC 249706694), except for the 16th dip in which we register the inner main belt asteroid 2009 BU179 at a distance of 36$''$.  Specifically, no known asteroid passes through the photometric aperture which fits within a box that extends only 20$''$ from the target star. Furthermore, on {\em Kepler} and {\em K2} the background flux was determined via a grid of background pixel targets to which a 2-D polynomial was fitted, and then evaluated at each pixel location \citep{twicken10}, so it is at best unlikely that asteroids could produce spurious transits unique to a particular target. And, in this latter regard, we note that while 28 such dips were recorded for EPIC 249706694, none was found in any of the 10 neighboring stars (see above discussion).

\section{Ground-Based Spectra}
\label{sec:spectra}

\subsection{TRES spectra}

We observed EPIC 249706694 (HD 139139) with the Tillinghast Reflector Echelle Spectrograph (TRES; \citealt{furesz08}) on the 1.5 m Tillinghast Reflector at the Fred Lawrence Whipple Observatory on Mt. Hopkins, Arizona. TRES has a spectral range of 3900-9100 Angstroms and a resolving power of R $\simeq$ 44,000. The spectra were reduced and extracted as described in \citet{buchhave10}.

We obtained two radial velocity observations on UT 2018 April 08 and April 10. The spectra had an average signal-to-noise per resolution element (SNRe) of $\simeq 35$  at the peak continuum near the Mg b triplet at 519 nm with exposure times averaging 240 seconds. A multi-order velocity analysis was performed by cross-correlating the two spectra, order by order. Twenty orders were used, excluding low S/N orders in the blue part of the spectrum and some red orders with contamination by telluric lines.  The difference in the relative velocity was about a few m s$^{-1}$, with an uncertainty of 36 m s$^{-1}$, based on the rms scatter of the differences from the individual orders.

The two radial velocities that we obtained with TRES are listed in Table \ref{tbl:RV}. Absolute radial velocities were derived for the two TRES spectra by cross-correlation against the best-matching synthetic template from the CfA library of calculated spectra using just the Mg b echelle order.  The two spectra, taken 2 nights apart, yield the same RV to within $4 \pm 100$ m s$^{-1}$. which is already quite constraining. 

The {\tt Stellar Parameter Classification} ({\tt SPC}; \citealt{buchhave10}) tool was used to derive the stellar parameters of the target star. {\tt SPC} cross correlates the observed spectra against a library of synthetic spectra based on Kurucz model atmospheres \citep{kurucz92}. We calculated the average of the inferred parameters for the two spectra, and they are fully consistent with those given in the EPIC input catalog \citep{huber16}.  The stellar parameters for Star A are summarized in Table \ref{tbl:spec}.  

\subsection{McDonald spectra}

We observed EPIC 249706694 with the high-resolution Tull spectrograph \citep{tull95} on the 2.7 meter telescope at McDonald Observatory in Ft.~Davis, TX.  We observed the target on 2018 April 6 using a 1.2 arcsecond wide slit, yielding a resolving power of 60,000 over the optical band. We obtained 3 individual spectra back-to-back with 20-minute exposures to aid in cosmic-ray rejection, which we combined in post-processing to yield a single higher signal-to-noise spectrum. We bracketed each set of 3 exposures with a calibration exposure of a ThAr arc lamp to determine the spectrograph's wavelength solution. We extracted the spectra from the raw images and determined wavelength solutions using standard IRAF routines, and we measured the star's absolute radial velocity using the {\tt Kea} software package \citep{endl_cochran16}. 

The McDonald spectrum of star A, taken 2 days earlier than the first TRES spectrum, agrees with the two TRES spectra to within $110 \pm 160$ m s$^{-1}$.  The stellar parameters for Star A independently inferred from the McDonald spectrum using {\tt Kea} are summarized in Table \ref{tbl:spec}.  

We also acquired three spectra of star B with the Tull spectrograph.  It was difficult, however, to completely avoid the light from the brighter nearby star A.  Nonetheless, we are able to extract RVs of $16.47 \pm 0.48$ km s$^{-1}$, $15.92 \pm 0.35$ km s$^{-1}$, and $16.06 \pm 0.46$ km s$^{-1}$ (see Table \ref{tbl:RV}). These values are quite consistent, within the uncertainties, with the RV for star A and show no evidence for variability, indicating that star B is likely not a short-period binary.    

\begin{table}
\centering
\caption{Radial Velocities}
\begin{tabular}{lccc}
\hline
\hline
Date & Radial Velocity & Star & Observatory \\
BJD-2450000 & km s$^{-1}$ & & \\
\hline
8216.8811  &   $16.558 \pm 0.100$  & A &  FLWO, TRES$^a$  \\  
8218.8975  &   $16.562 \pm 0.100$  & A &  FLWO, TRES$^a$ \\   
8214.8914    &   $16.45 \pm 0.14$    & A &  McDonald \\
6863--7562        &   $16.36 \pm 0.40$   & A &  Gaia$^b$ \\
2809  &  $19.7 \pm 4.6$ & A & RAVE$^c$ \\ 
4598  &   $17.7 \pm 1.1$ & A & RAVE$^c$ \\ 
\hline
8261.8914    &   $16.47 \pm 0.48$   & B  & McDonald \\ 
8587.9312    &   $15.92 \pm 0.35$   & B  & McDonald \\  
8602.9276    &   $16.06 \pm 0.46$   & B  & McDonald \\   
6863--7562   &   $15.54 \pm 3.97$   & B  &  Gaia$^b$ \\
\hline
\label{tbl:RV}
\end{tabular}  

{\bf Notes.} (a) We have applied a zero-point offset of $-0.61$ km s$^{-1}$ to match the IAU RV standard stars corrected to the barycenter.  (b) The cited RV is the median of the values obtained over 7 visits and we therefore list only the range of times when the DR2 data were acquired.  The error bar is the standard error deduced from the 7 measurements (see Sect.~\ref{sec:gaia}).  (c) Heliocentric values, and we therefore give only the observation time to the nearest day (see Sect.~\ref{sec:rave}).  
\end{table}

\subsection{Gaia spectra}
\label{sec:gaia}

The Gaia data base \citep{lindegren18} lists an RV value for star A of $16.36 \pm 0.40$ km s$^{-1}$ which agrees with the two TRES spectra to within $200 \pm 400$ m s$^{-1}$.   It also turns out that Gaia was able to get a rough radial velocity for star B which differs from that of star A by only $1 \pm 4$ km s$^{-1}$. We note, however, that the reported Gaia RV values are the median RVs for 7 and 6 visits for star A and star B, respectively, over a $\sim$2 year interval.  The assigned uncertainty is roughly the standard error of the set of measurements\footnote{ For details see pp.~531-532 of the DR2 documentation \url{https://gea.esac.esa.int/archive/documentation/GDR2/pdf/GaiaDR2_documentation_1.1.pdf}.}.  Thus, the upper limit to the K velocity of the binary motion for any orbital period that we might consider (e.g., $P_{\rm orb} \lesssim$ yr) is less than $\sim$ km s$^{-1}$ for star A and 10 km s$^{-1}$ for star B, based on the Gaia data alone.   

\subsection{RAVE spectra} 
\label{sec:rave}

 The 5th data release of the RAdial Velocity Experiment (`RAVE', \citealt{kunder17})\footnote{\url{https://www.rave-survey.org/project/documentation/dr5/}} lists two RV measurements for star A taken in 2003 and 2008.  They report heliocentric velocities of $19.7 \pm 4.6$ and $17.7 \pm 1.1$ km s$^{-1}$ for the two measurements, respectively.  These are also listed in Table \ref{tbl:RV}.  Though, not highly precise, these also indicate long-term stability in the radial velocity of star A.

\begin{table}
\centering
\caption{Spectral Properties}
\begin{tabular}{lccc}
\hline
\hline
Parameter & Value & Star & Observatory \\
\hline
$T_{\rm eff}$ [K]  &   $5820 \pm 50$  & A &  FLWO, TRES  \\
$T_{\rm eff}$ [K]  &   $5875 \pm 125$  & A &  McDonald \\
$\log g$ [cgs]    &   $4.35 \pm 0.10$    & A &  FLWO, TRES \\
$\log g$ [cgs]    &   $4.50 \pm 0.12$   & A &  McDonald \\
$v \sin i$ [km/s] &   $3.7 \pm 0.5$    & A &  FLWO, TRES \\
$v \sin i$ [km/s] &   $3.2 \pm 0.3$    & A &  McDonald \\
$[{\rm m/H}]$ &   $0.05 \pm 0.08$    & A &  FLWO, TRES \\
$[{\rm m/H}]$  &  $0.0 \pm 0.1$  & A &  McDonald \\
\hline
$T_{\rm eff}$ [K]  &   $4500 \pm 100$  & B &  McDonald \\   
$\log g$ [cgs]    &   $4.2 \pm 0.4$   & B &  McDonald \\  
$v \sin i$ [km/s] &   $3.4 \pm 0.4$    & B &  McDonald \\  
\hline
\label{tbl:spec}
\end{tabular}  

{\bf }   
\end{table}

\subsection{Summary of spectral studies}

From all the spectral observations discussed above we draw two conclusions:  (i) stars A and B likely form a bound binary (projected separation $\sim$400 AU) inferred from the similar RVs and distances, comparable proper motions, and close proximity on the sky; and (ii) neither star is itself in a close binary system because of lack of large RV variations.   

We can further quantify the latter conclusion for star A.  We utilized all the RV values listed in Table \ref{tbl:RV} to carry out a Fisher-matrix analysis (D.~Wittman\footnote{\url{wittman.physics.ucdavis.edu/Fisher-matrix-guide.pdf}}) of the (2-$\sigma$) upper limits to circular orbits with nearly $10^6$ trial periods between 0.5 d and 30 years.  We then adopted a mass for star A of 1.0 $M_\odot$ and considered inclination angles from 30$^\circ$ to 90$^\circ$ in order to set corresponding limits on the mass of any unseen close companion to star A.  The result is that stars and even brown dwarfs are ruled out for $P_{\rm orb} \lesssim 10$ d, low-mass stellar companions are disallowed for $P_{\rm orb} \lesssim 100$ d, and comparable mass stars are ruled out for periods below a few decades. 

\section{Possible Explanations for the Dips}
\label{sec:scenarios}

\subsection{Planet transits in a multiplanet system} 
\label{sec:41}

We have shown that no more than subsets of 4 dips could be caused by periodic transits.  Let's assume for the sake of argument that the host star has a chain of planets in 3:2 mean motion resonances. If the first has a period of 20 days, yielding 4 visible transits during C15, then the next planet could yield at most 3 transits, and so forth.  Adding these up, we would see at most 13 transits before longer period planets began yielding, at best, a single improbable transit. It is farfetched to imagine that each of the remaining 15 transits could be accounted for by different individual planets, each more distant than the previous one, and still yielding transits all within an 87-day observation interval.

\subsection{Planets orbiting both stars A and B}
\label{sec:42}

This hypothesis does not help much over that described in Sect.~\ref{sec:41}.  It is also farfetched to imagine that each of star A and star B has a subset of the transits (e.g., 14 each), and only small subsets of each of these would be periodic---yet all of the transit depths are similar.  This, in spite of the fact that the putative planets around star A would all have to be Earth-size while all around star B would be Jovian size.   

\subsection{Few-planet system with huge TTVs}
\label{sec:43}

We also considered a case where the transits have sufficiently large TTVs so as not to be detectable in BLS searches.  In that case, they should still be detectable in the IMT transform that we computed.  In fact, we allowed TTVs as large as plus and minus 10\% of a period, and the IMT still did not find any underlying periodicity.  We cannot envision a scenario where huge TTVs, e.g., causing a pattern such as seen in Fig.~\ref{fig:rawLC}, can exist, and where the system remains dynamically stable.  

\subsection{Disintegrating planet with only rare transits appearing}
\label{sec:44}

Another possibility is an ultrashort period planet (`USP') where the transits are caused by dusty effluents rather than a hard body occultation (e.g., \citealt{rappaport12}; \citealt{vanlieshout18}). In that case, perhaps only a few of the transits would appear.  Since the shortest intervals between `transits' in the observed data set are $\sim$5 hours, this would have to be the maximum allowed period of the putative USP.  In that case, there would have been more than 400 transits during the C15 observations.  If only 28 of them are observed, then one would need to explain why the dust emissions were sufficiently active in only 7\% of the events to reveal a transit, but nearly completely absent in the remaining 93\% of cases.  In any event, we have done simulations to show that a random selection of 10-15\% of several hundred periodic events is readily detectable in a BLS transform, but 7\% is getting close to undetectable.  Finally, in this regard, we note that the average dip profile (see Fig.~\ref{fig:stack}) shows no evidence for the type of asymmetries that might be expected from disintegrating planets.

\subsection{Debris disk of dust-emitting asteroids}
\label{sec:45}

An initially attractive scenario is a debris disk hosting many dust-emitting asteroids or planetesimals over a substantial range of radial distances.  Each transit-like event might be attributed to a single asteroid with dusty effluents transiting the host star.  We note that the postulated dust-emitting asteroids in WD 1145+017 \citep{vanderburg15} could produce 200 ppm transits when crossing a Sun-like star.  The difficulties with such a scenario include: (i) the fact that almost all the transit-like events have rather similar depths, and (ii) it is doubtful that asteroids in a wide range of orbits would all be at just the right equilibrium temperatures to produce copious dust emission.  These both point to the problem with numerous independent bodies all having just the right properties to emit very nearly the same amounts of dust.

\subsection{Eccentric planet orbit about one star in a close binary -- S-type}
\label{sec:46}

One interesting possibility is that of a planet in an eccentric orbit about one star in a fairly close binary system, e.g., with a binary period of 2--5 days.  The eccentric planetary orbit would require a very short orbital period of $\lesssim 5-16$ hours to remain well inside the Roche lobe of the host star.  The rapid precession of such an eccentric planetary orbit would allow for both different durations of the transits and a possible erratic set of arrival times.  The putative binary system would have to be sufficiently tilted with respect to our line of sight so as to avoid producing binary eclipses -- which are not observed.  We have carried out a number of such dynamical simulations and found that we could never succeed in masking the underlying periodicity sufficiently to avoid detection in either a BLS or IMT transform. Furthermore, the radial velocity information we have (Table \ref{tbl:RV}) also argues against a short-period binary, as does the fact that there is no sign of a composite spectrum in the TRES spectral observations.

\subsection{P-type circumbinary planet}
\label{sec:47}

Another scenario we investigated is a P-type circumbinary planet. Transits across one or both stars and the stellar motions around the center of mass may produce large variations in the transit timing and duration.  This would be an extreme case of transiting circumbinary planets found by \textit{Kepler} \citep[e.g.,][]{doyle11, welsh12}.
To test this scenario, we performed the following simple analysis. We initialize the planetary orbit such that it eclipses one of the binary stars when the first transit event occurs at BKJD = 3161.6826. We ignored dynamical interactions in this simple model, i.e., we assumed that the binary and planetary motions follow independent Keplerian orbits with period ratios outside the stability limit \citep{holman99}. To sample the orbital parameters, we define a likelihood function that is the sum of booleans of all individual {\em K2} data points: (a) true if the time is within one of the transiting events (Table 1) and the planet is in front of one of the stars from the observer's view in our model; (b) false if a transit is observed but the planet is not eclipsing in our model, or the planet eclipses in our model but no transit was observed.  We minimized the corresponding likelihood function with a Levenberg-Marquardt algorithm to find the best solutions. However, after initializing the model with various starting parameters ($P_{\rm orb}$ typically near 2 d and $P_{\rm plan}$ near 8-10 d), we could not find a convincing solution that correctly predicted more than $\sim$5 of the events in Table 1. Given the crudeness of our model, the P-type circumbinary scenario is perhaps more decisively ruled out by the absence of RV variations in the host star (see Sect.~\ref{sec:spectra}). 

\subsection{Dipper-like activity}
\label{sec:48}

Some `dipper' stars \citep{cody14,ansdell16} exhibit quasi random dips that often range from a few to 50\% of the flux. If such dipper activity happens to take place in association with the fainter star B, then those dips would be diluted by about a factor of 15 due to the presence of star A which is in the same photometric aperture.  To check the plausibility of this scenario, we have taken the lightcurves for the 10 dipper stars reported in \citet{ansdell16} (see their Fig.~2), and have mocked up what one would have seen if the variability in them had been diluted by a factor of 15.  In no case does the resultant `diluted' version of the lightcurves closely resemble the lightcurve of the Random Transiter.  In particular, at least one of the following is likely to apply: (i) there is significant activity in the out-of-`dip' regions; or (ii) the host star has excess WISE band 3 \& 4 emissions; or (iii) there is an underlying periodicity present comparable with the dip recurrence times.  The fact that these are all different from the Random Transiter, and star A in particular, does not prove that  `dipper' phenomena are not involved in the transit-like events we see.  For one example, star B might have a small WISE band 3 \& 4 excess that we are unable to detect because of the dilution of star A. Finally, we note that since there is currently no quantitative model for explaining dipper stars, invoking dipper behavior to explain the Random Transiter does not provide any real insight to the mystery at hand. 

\subsection{Short-lived star spot}
\label{sec:49}

Finally, we considered cases where the dips in flux are caused by a short-lived change in the intrinsic flux from the host star.  One idea is that the host star has short-lived spots that are larger than the Earth in size on star A, or larger than Jupiter on star B, but appear suddenly (i.e., within a half hour), last for a few hours, and then disappear equally suddenly.  We do not believe that this violates any physical laws.  Furthermore, the existence of such spot activity on distant stars studied with {\em Kepler}, {\em K2}, and {\em TESS}, might have gone either unnoticed or unreported by Citizen Scientists or by automated searches for single-transit systems.  This is especially true if such dips are much less frequent than once per month, are aperiodic, are at the 200 ppm level, and last for only a couple of data cadences. In the latter case, they might be considered too short for the implied longer periods associated with single-transit events.    

\section{Summary and Conclusions}
\label{sec:summary}

K2 has provided us with a large number of impressive results\footnote{\url{https://keplerscience.arc.nasa.gov/scicon-2019/}}.  Several mysterious objects, however, linger even after the end of the mission.  In this work we have presented a {\em K2} target, EPIC 249706694 (HD 139139), that exhibits 28 transit-like events that appear to occur randomly.  The depths, durations, and times of these dips are shown in Figures~\ref{fig:rawLC} and \ref{fig:DD_corr} and summarized in Table \ref{tbl:transits}.  Most of the events have depths of $200 \pm 80$ ppm, durations of 1-7 hr, and a mean separation time of $\sim$3 days.

The target star EPIC 249706694 (star A) is a relatively unevolved star of spectral type G.  It has a rotation period of $\simeq 14.5$ days, which implies a gyrochronological age of $1.5 \pm 0.4$ Gyr \citep{barnes07}.  However, in the same photometric aperture is a fainter neighbouring star (3.3$''$ away) that is 2.7 $G_{\rm rp}$ magnitudes fainter with $T_{\rm eff} \simeq 4400$ K.  There is insufficient Gaia kinematic data for this star (B) to know for sure if it is bound to star A, but its similar RV value, distance, and estimated close relative proper motion (see Sect.~\ref{sec:archive}) imply that it is likely a physically bound companion to star A.  In any case, we do not know with certainty which star hosts the transit-like events.  If the dips are transits across star A then the inferred planet sizes would be $\simeq 2\,R_\oplus$, whereas if they are on star B, then the planetary radii would be more like 1-2 times Jupiter's size.

Neither star A nor star B appears to exhibit large RV variations, with limits of $\lesssim 1$ km s$^{-1}$ (see Sect.~\ref{sec:spectra} and Table \ref{tbl:RV}). The constraints that the RV measurements place on possible orbital periods for binary companions are discussed at the end of Sect.~\ref{sec:spectra}.  For star A we find that stars and even brown dwarfs are ruled out for $P_{\rm orb} \lesssim 10$ d, and low-mass stellar companions are disallowed for $P_{\rm orb} \lesssim 100$ d. For star B the results are somewhat less constraining.

We have done some exhaustive tests to ensure that the observed shallow dips in the flux of EPIC 249706694 are of astrophysical origin (see Sect.~\ref{sec:tests}). These include ruling out rolling-band artefacts, centroid motions during the dips, cross-talk, or bad CCD pixels.  One can never be entirely sure that such transit-like events are astrophysical, but we have done our best due diligence in this regard.

We have briefly considered a number of scenarios for what might cause the randomly occurring transit-like events.  These include actual planet transits due to multiple or dust emitting planets, dust-emitting asteroids, one or more planets with huge TTVs, S- and P-type transits in binary systems, `dipper' activity, and short-lived spot activity.  We find that none of these, though intriguing, is entirely satisfactory. These are reviewed in Sect.~\ref{sec:scenarios}.

The purpose of this paper is largely to bring this enigmatic object to the attention of the larger astrophysics community in the hope that (i) some time on larger telescopes, or ones with high photometric precision, might be devoted to its study,  and (ii) some new ideas might be generated to explain the mysterious dips in flux.  With regard to observational advances, two specific things could be done.  First, acquire more and higher quality spectra of stars A and B to improve the constraints on RV variability.  Second, if time is available on a telescope with 1 mmag precision photometry in the presence of a brighter star 3.3$''$ away, monitor star B for a total exposure time exceeding a few days.  There would then be a chance of directly measuring the transit-like events on star B -- if that is indeed their origin.  

\vspace{20pt}

\noindent
{\bf Acknowledgements} 

A.\,V.'s and K.\,M.'s work was supported in part under a contract with the California Institute of Technology (Caltech)/Jet Propulsion Laboratory (JPL) funded by NASA through the Sagan Fellowship Program executed by the NASA Exoplanet Science Institute.
We thank Allan R. Schmitt and Troy Winarski for making their lightcurve examining software tools {\tt LcTools} and {\tt AKO-TPF} freely available.  
Some of the data presented in this paper were obtained from the Mikulski Archive for Space Telescopes (MAST). STScI is operated by the Association of Universities for Research in Astronomy, Inc., under NASA contract NAS5-26555. Support for MAST for non-HST data is provided by the NASA Office of Space Science via grant NNX09AF08G and by other grants and contracts. 
This research has made use of IMCCE's SkyBoT VO tool.



\begin{thebibliography}{}

\bibitem[Ansdell et al.(2016)]{ansdell16} Ansdell, M., Gaidos, E., Rappaport, S., et al. 2016, ApJ, 816, 69

\bibitem[Bailer-Jones et al.(2018)]{bailerjones18} Bailer-Jones, C.A.L., Rybizki, J., Fouesneau, M., Mantelet, G., \& Andrae, R. 2018, AJ, 156, 58

\bibitem[Barentsen \& Cardoso (2018)]{barensten18} Barensten, G., \& Cardosos, J.V.d.M. 2018, Astrophysics Source Code Library, ascl.soft03005B

\bibitem[Barnes (2007)]{barnes07} Barnes, S.A. 2007, ApJ, 669, 1167

\bibitem[Becker et al.(2015)]{becker15} Becker, J.C., Vanderburg, A., Adams, F.C., Rappaport, S., \& Schwengeler, H.M. 2015, ApJL, 812, L18

\bibitem[Berthier et al.(2006)]{berthier6} Berthier, J., Vachier, F., Thuillot, W., Fernique, P., Ochsenbein, F., Genova, F., Lainey, V., \& Arlot, J.-E. 2006, ASPC, 351, 367

\bibitem[Berthier et al.(2016)]{berthier16}  Berthier, J., Carry, B., Vachier, F., Eggl, S., \& Santerne, A. 2016, MNRAS, 458, 3394

\bibitem[Borkovits et al.(2018)]{borkovits18} Borkovits, T., Albrecht, S., Rappaport, S., et al. 2018, MNRAS, 478, 5135

\bibitem[Borkovits et al.(2019)]{borkovits19} Borkovits, T., Rappaport, S., Kaye, T., et al. 2019, MNRAS, 483, 1934

\bibitem[Buchhave et al.(2010)]{buchhave10} Buchhave, L.A., Bakos, G.\'A., Hartman, J.D. et al. 2010, ApJ, 720,1 118

\bibitem[Carter et al.(2011)]{carter11} Carter, J.A., Fabrycky, D.C., Ragozzine, D., et al. 2011, Sci, 331, 562

\bibitem[Cash (1979)]{cash79} Cash, W. 1979, ApJ, 228, 939

\bibitem[Chaplin et al.(2015)]{chaplin15} Chaplin, W.J., Lund, M.N., Handberg, R., et al. 2015, PASP, 127, 1038

\bibitem[Christiansen et al.(2013)]{christiansen13} Christiansen, J.L., Clarke, B.D., Burke, C.J., et al. 2013, ApJS, 207, 35

\bibitem[Cody et al.(2014)]{cody14} Cody, A.M., Stauffer, J., Baglin, A., et al. 2014, ApJ, 147, 82

\bibitem[Cutri et al.(2003)]{cutri03} {Cutri}, R.~M., {Skrutskie}, M.~F., {van Dyk}, S., et al. 2003, The IRSA 2MASS All-Sky Point Source Catalog, NASA/IPAC Infrared Science Archive

\bibitem[Cutri et al.(2013)]{cutri13} Cutri, R.M., Wright, E.L., Conrow, T., et al.~2013, wise.rept, 1C.

\bibitem[Dattilo et al.(2019)]{dattilo19}Dattilo, A., Vanderburg, A., Shallue, C.J., et al. 2019, AJ, 157, 169

\bibitem[Dimitriadis et al.(2019)]{dimitriadis19} Dimitriadis, G., Foley, R.J., Rest, A., et al. 2019, ApJL, 870,1

\bibitem[{Doyle} {et~al.}(2011)]{doyle11} {Doyle}, L.~R., {Carter}, J.~A., {Fabrycky}, D.~C., {et~al.} 2011, Science, 333, 1602

\bibitem[Endl \& Cochran (2016)]{endl_cochran16} Endl, M., \& Cochran, W.D. 2016, PASP, 128, 94502

\bibitem[F\H ur\'esz et al.(2008)]{furesz08}F\H ur\'esz, G., Szentgyorgyi, A. H., \& Meibom, S. 2008, in {\em Precision Spectroscopy in Astrophysics}, ed. N. C. Santos, L. Pasquini, A. C. M. Correia, \& M. Romaniello, 287, 290

\bibitem[Flewelling et al.(2016)]{flewelling16} Flewelling, H.A., Magnier, E.A., Chambers, K.C., et al. 2006, ``The Pan-STARRS1 Database and Data Products'', arXiv:1612.05243

\bibitem[Gaidos et al.(2014)]{gaidos14} Gaidos, E., Mann, A.W., L\'epine, S., et al. 2014, MNRAS, 443, 2561

\bibitem[Garnavich (2019)]{garnavich19}  Garnavich, P. 2019, ``Better Understanding Supernovae from {\em Kepler/K2} Observations'', talk presented at the Fifth {\em Kepler} \& {\em K2}; {\em Kepler} Science Center Website, DOI:10.5281/zenodo.593417

\bibitem[Gary et al.(2017)]{gary17} Gary, B.L., Rappaport, S., Kaye, T.G., Alonso, R., Hambschs, F.-J. 2017, MNRAS, 465, 3267

\bibitem[Grindlay et al.(2009)]{grindlay09} Grindlay, J., Tang, S., Simcoe, R., Laycock, S., et al. 2009, in: {\it Preserving Astronomy's Photographic Legacy: Current State and the Future of North American Astronomical Plates}, ed. W. Osborn \& L. Robbins, (San Francisco, CA), ASP Conf. Ser. 410, 101

\bibitem[Holman \& Wiegert (1999)]{holman99} Holman, M.J., \& Wiegert, P.A. 1999, AJ, 117, 621

\bibitem[Howell et al.(2014)]{howell14} Howell, S.B., Sobeck, C., Hass, M., et al. 2014, PASP, 126, 398

\bibitem[Huber et al.(2016)]{huber16} Huber, D., Bryson, S.T., Haas, M.R., et al. 2016, ApJS, 224, 2

\bibitem[Jayasinghe et al.(2018)]{jayasinghe18} Jayasinghe, T., Kochanek, C.S., Stanek, K.Z., et al. 2018, MNRAS, 477, 3145

\bibitem[Jenkins et al.(2010)]{jenkins10} Jenkins, J.M., Caldwell, D.A., Chandrasekaran, H., et al. 2013, ApJL, 713, L87

\bibitem[Kipping et al.(2015)]{kipping15} Kipping, D. M., Schmitt, A. R., Huang, X., Torres, G., Nesvorn\'y, D., Buchhave, L. A., Hartman, J., \& Bakos, G. \'A. 2015,  ApJ, 813, 14

\bibitem[Koch et al.(2010)]{koch10} Koch, D.G., Borucki, W.J., Basri, G., et al. 2010, ApJL, 713, L79

\bibitem[Kov\'acs et al.(2002)]{kovacs02}  Kov\'acs, G., Zucker, S., \& Mazeh, T. 2002, A\&A, 391, 369

\bibitem[{Kunder} {et~al.}(2017)]{kunder17} {Kunder}, A., {Kordopatis}, G., {Steinmetz}, M., {et~al.} 2017, AJ, 153, 75

\bibitem[Kurucz (1992)]{kurucz92}Kurucz, R. L. 1992, in: IAU Symposium, Vol. 149, {\it The Stellar Populations of Galaxies}, ed. B. Barbuy \& A. Renzini, 225

\bibitem[Leiner et al.(2019)]{leiner19} Leiner, E., Mathieu, R.D., Vanderburg, A., Gosnell, N.M., \& Smith, J.C. 2019, arXiv:1904.02169

\bibitem[Lindegren et al.(2018)]{lindegren18} Lindegren, L., Hernandez, J, Bombrun, A., et al. 2018, arXiv:1804.09366.

\bibitem[Luger et al.(2016)]{luger16} Luger, R., Agol, E., Kruse, E., Barnes, R., Becker, A., Foreman-Mackey, D., \& Deming, D. 2016, AJ, 152, 100L

\bibitem[Lund et al.(2016a)]{lund16a} Lund, M.N., Chaplin, W.J., Casagrande, L., et al. 2016a, PASP, 128, 124204

\bibitem[Lund et al.(2016b)]{lund16b} Lund, M.N., Basu, S., Silva Aguirre, V., et al. 2016b, MNRAS, 463, 2600 

\bibitem[Malavolta et al.(2018)]{malavolta18} Malavolta, L., Mayo, A.M., Louden, T., et al. 2018, AJ, 155, 107

\bibitem[Mayo et al.(2018)]{mayo18} Mayo, A.W., Vanderburg, A., Latham, D., et al. 2018, AJ, 155, 136
    
\bibitem[Moln\'ar et al.(2018)]{molnar18} Moln\'ar, L., P\'al, A., S\'arneczky, K., et al. 2018, ApJS, 234, 37

\bibitem[Rappaport et al.(2012)]{rappaport12} Rappaport, S.A., Levine, A., Chiang, E., et al. 2012, ApJ, 752, 1

\bibitem[Rappaport et al.(2017)]{rappaport17} Rappaport, S., Vanderburg, A., Borkovits, T., et al. 2017, MNRAS, 467, 2160

\bibitem[Rappaport et al.(2019)]{rappaport19} Rappaport, S., Zhou, G., Vanderburg, A., et al. 2019, MNRAS, 485, 2681

\bibitem[Rest et al.(2018)]{rest18} Rest, A., Garnavich, P.M., Khatami, D., et al. 2018, NatAs, 2, 307

\bibitem[Ricker et al.(2014)]{ricker14} Ricker, G., Winn, J.N., Vanderspek, R., et al. 2014, SPIE, 9143, 20

\bibitem[Rodriguez et al.(2018)]{rodriquez18} Rodriguez, J.E., Becker, J.C., Eastman, J.D., et al. 2018, AJ, 156

\bibitem[Sanchis-Ojeda et al.(2016)]{sanchisojeda15} Sanchis-Ojeda, R., Rappaport, S., Pall`e, E., et al. 2015, ApJ, 812, 112

\bibitem[Skrutskie et al.(2006)]{skrutskie06} Skrutskie, M.F., Cutri, R.M., Stiening, R., et al. 2006, AJ, 131, 1163.

\bibitem[Smart \& Nicastro(2014)]{smart14} Smart, R.L., \& Nicastro, L. 2014, \aap, 570, 87

\bibitem[Stello et al.(2017)]{stello17} Stello, D., Zinn, J., Elsworth, Y., et al. 2017, ApJ, 835, 83

\bibitem[Thompson et al.(2016)]{thompson16} Thompson, S.E., Fraquelli, D., van Cleve, J.E., \& Caldwell, D.A. 2016, Kepler Archive Manual (KDMC-10008-006) 

\bibitem[Tovmassian et al.(2018)]{tovmassian18} Tovmassian, G., Szkody, P., Yarza, R., \& Kennedy, M.  2018, ApJ, 863, 47

\bibitem[Tull et al.(1995)]{tull95} Tull, R.G., MacQueen, P.J., Sneden, C., \& Lambert, D.L. 1995, PASP, 107, 251

\bibitem[Twicken et al.(2010)]{twicken10} Twicken, J.D., Clarke, B.D., Bryson, S.T., et al.  2010, Software and Cyberinfrastructure for Astronomy, 774023

\bibitem[Van Cleve \& Caldwell(2016)]{vancleve16} Van Cleve, J.E., \& Caldwell, D.A.\ 2016, Kepler Science Document, KSCI-19033-002, Edited by Michael R.~Haas and Steve B.~Howell

\bibitem[Vanderburg \& Johnson(2014)]{vanderburg14} Vanderburg, A., \& Johnson, J.A. 2014, PASP, 126, 948

\bibitem[Vanderburg et al.(2015)]{vanderburg15} Vanderburg, A., Johnson, J.A., Rappaport, S. 2015, Nature, 526, 546

\bibitem[Vanderburg et al.(2016)]{vanderburg16} Vanderburg, A., Becker, J.C., Krisitansen, M.H., et al. 2016, ApJL, 827, 10

\bibitem[van Lieshout \& Rappaport (2018)]{vanlieshout18} van Lieshout, R. \& Rappaport, S. 2018, arXiv:1708.00633, Handbook of Exoplanets, Edited by Hans J. Deeg and Juan Antonio Belmonte. Springer Living Reference Work, ISBN: 978-3-319-30648-3, 2017, id.15

\bibitem[{Welsh} {et~al.}(2012)]{welsh12} {Welsh}, W.~F., {Orosz}, J.~A., {Carter}, J.~A., {et~al.} 2012, Nature, 481, 475

\bibitem[Zhou et al.(2018)]{zhou18} Zhou, G., Rappaport, S., Nelson, L., et al. 2018, \apj, 854, 109

\bibitem[Zhu et al.(2017)]{zhu17} Zhu, W., Udalski, A., Huang, C.X., et al. 2017, ApJL, 849, 31

\end{thebibliography}
\end{document}